\documentclass[prb,aps,twocolumn,amsmath,amssymb,superscriptaddress]{revtex4}

\usepackage[dvips]{epsfig}
\usepackage{float}
\usepackage{tabularx}
\usepackage{makecell}
\usepackage{color}
\usepackage[sort&compress]{natbib}
\usepackage{helvet}
\makeatletter
\renewcommand{\@biblabel}[1]{#1. }
\renewcommand{\@dotsep}{500}
\renewcommand{\@pnumwidth}{0em}
\renewcommand{\l@figure}[2]{
        \@dottedtocline{1}{1.5em}{2em}{Figure #1}{}\vspace{15pt}}

\newcommand*{\citen}[1]{%
  \begingroup
    \romannumeral-`\x 
    \setcitestyle{numbers}%
    \cite{#1}%
  \endgroup   
}

\begin{document}
\title{Entangled photon-pair generation\\ in periodically-poled thin-film lithium niobate waveguides}
\author{Jie Zhao}
\affiliation{University of California, San Diego, Department of Electrical and Computer Engineering, La~Jolla, California 92093-0407, USA}
\author{Chaoxuan Ma}
\affiliation{University of California, San Diego, Department of Electrical and Computer Engineering, La~Jolla, California 92093-0407, USA}
\author{Michael R{\"u}sing}
\affiliation{University of California, San Diego, Department of Electrical and Computer Engineering, La~Jolla, California 92093-0407, USA}
\affiliation{TU Dresden, Institut f{\"u}r Angewandte Physik, N{\"o}thnitzer Stra{\ss}e 61, 01187 Dresden, Germany}
\author{Shayan Mookherjea}
\affiliation{University of California, San Diego, Department of Electrical and Computer Engineering, La~Jolla, California 92093-0407, USA}
\email{smookherjea@ucsd.edu}
\date{\today}
\pacs{}

\begin{abstract}
We report measurements of time-frequency entangled photon pairs and heralded single photons at telecommunications wavelengths, generated using a periodically-poled, lithium niobate on insulator (LNOI) waveguide pumped optically by a diode laser. We achieve a high Coincidences-to-Accidentals Ratio (CAR) at high pair brightness, a low value of the conditional self-correlation function [$g^{(2)}(0)$], and high two-photon energy-time {F}ranson interferometric visibility, which demonstrate the high quality of the entangled photon pairs and heralded single photons.     
\end{abstract}

\maketitle

\section{Introduction}
Integrated photonics can be useful in generating, manipulating and detecting non-classical light, including photon pairs and heralded single photons as resources for quantum optical communications and information processing. Compared to bulk nonlinear optical crystals, the use of waveguides and periodic poling has led to significant improvements in brightness, quality and simplicity of near-infrared wavelength photon pair sources.\cite{Alibart2016,inbook-Silberhorn} For photon pair generation using nonlinear optical processes (e.g., spontaneous parametric down conversion, SPDC, or spontaneous four-wave mixing, SFWM), the intrinsic rate of nonlinear optical processes increases as the cross-sectional area of the waveguide mode decreases. Generally, sub-micron modal area (i.e., $A_\text{eff} \le 1\ \mathrm{\mu m}^2$) waveguides are associated with high-index contrast silicon photonics, in which SFWM generates a reasonably high rate of photon pair generation at sub-milliwatt pump power levels.\cite{Sharping2006}

The objective of this paper is to report good performance achieved using SPDC in a periodically-poled thin-film lithium niobate (LN) waveguide. The cross section and top-down view of the lithium niobate on insulator (LNOI) waveguide, which supports well-defined, quasi-TE polarized fundamental waveguide modes with sub-micron $A_\text{eff}$ is shown in Fig.~1. The performance of LNOI based devices can be superior to that of traditional LN waveguide devices; one notable example is that of $>$100~GHz bandwidth electro-optic modulators.~\cite{Wang2018a,Weigel2018} However, the reported SPDC performance of LNOI~\cite{Frank2016,Luo2017,Rao2018a,Chen2019a_YPH_arxiv} in terms of the usual metrics such as Coincidences-to-Accidentals Ratio (CAR) and conditional self-correlation [$g^{(2)}(0)$] has yet to catch up to SPDC in traditional LN waveguides, where coincidences-to-accidentals ratio $\text{CAR} > 10,000$, heralded two-photon auto-correlation $g^{(2)}(0) < 0.01$, and two-photon interferometric visibility $V\approx 99\%$ are common.\cite{art-Brida-2012,Krapick2013,Harder:13,Inagaki:13,Bock:16} 

In section~\ref{sec-design}, we present our waveguide design and describe the poling procedure, along with diagnostic images of the poling. We have recently developed a high-quality poling recipe for x-cut LNOI, and demonstrated two useful diagnostic methods using in-situ poling monitoring, and a non-destructive nonlinear microscopy technique.\cite{Zhao2019,Ruesing2019} This allows us to improve the poling process, and identify suitable waveguides without destroying them e.g., by cross-sectioning, and etching using dilute acids, as is typically performed in traditional poling diagnostics. In section~\ref{sec-experiment}, we describe our experiments in performing SPDC, where we used a continuous-wave optical pump at 784.5 nm in order to generate frequency-degenerate photon pairs at 1569 nm. In sections \ref{subsec-CAR}, \ref{subsec-g2} and \ref{subsec-Entanglement}, we describe our results in frequency-degenerate photon-pair generation, heralding of single photons from the pair, and verification of time-energy entanglement, respectively, which are basic resources in quantum optics technology. Here, we show that these waveguides overcome the previously-mentioned performance gap with traditional PPLN waveguides for basic SPDC, and, in fact, demonstrate a large improvement over traditional waveguides in achieving high CAR at high PGR.  

\section{Experimental details: waveguide design}
\label{sec-design}

\begin{figure*}[tbh]
\centering
\includegraphics[width=\linewidth, clip=true]{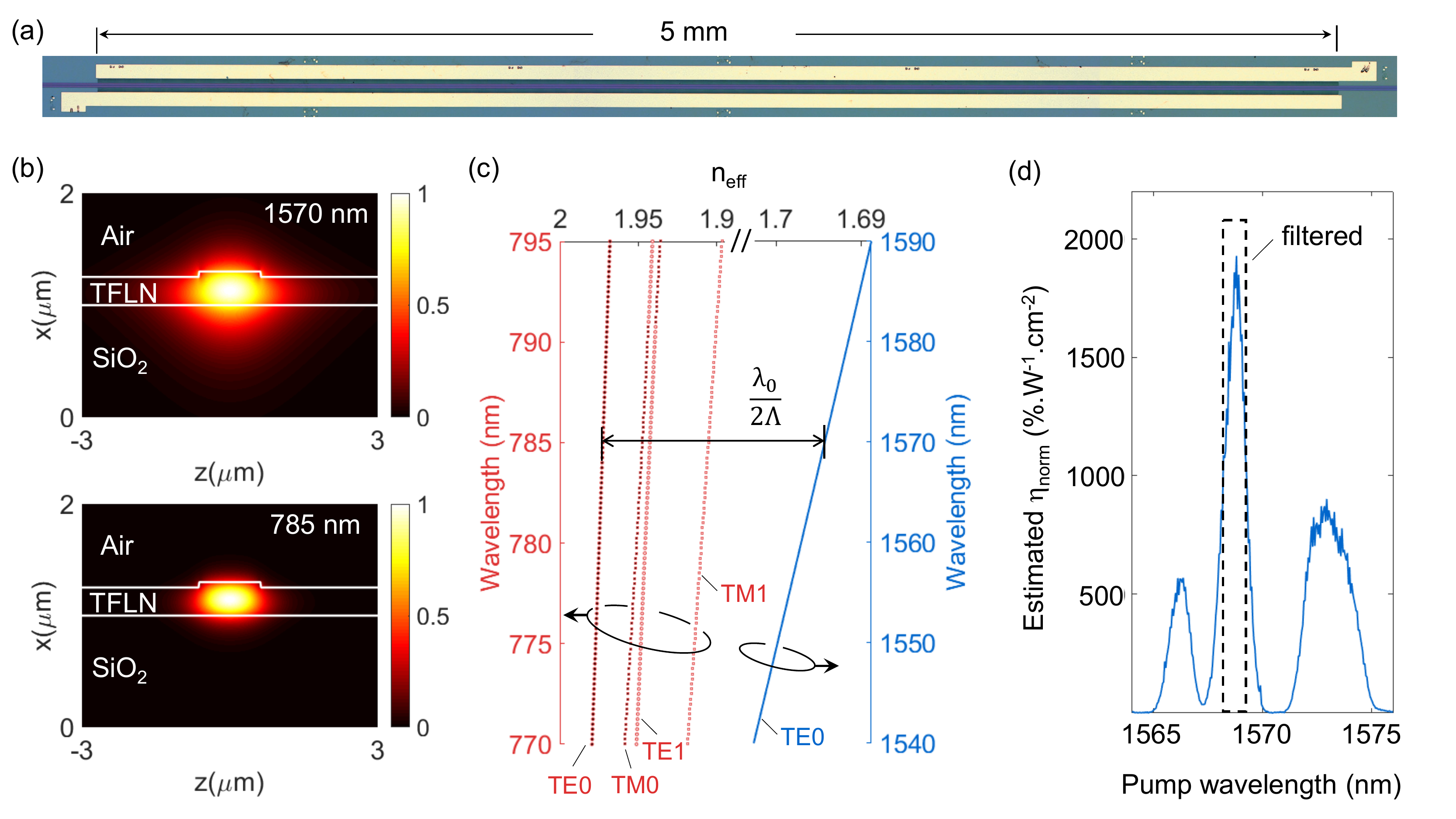}
\caption{(a) Optical microscope image of a fabricated waveguide (7~mm long) with a poled region between electrodes (5~mm long). (b) Calculated profile (magnitude of the major electric-field component) of the quasi-TE-polarized fundamental mode at 1570~nm for the signal and idler photons. (b) Calculated profile of the quasi-TE-polarized mode at 785 nm for the pump of the SPDC process. (c) Effective refractive indices with the criterion for type-0 quasi phase matching indicated by the horizontal, double-headed arrow. (d) Measured efficiency of second harmonic generation (SHG), from wavelengths near 1570 nm to wavelengths near 785 nm.} 
\label{fig-1}
\end{figure*}

The devices were designed and fabricated in a 300 nm thickness MgO-doped x-cut LN thin film, which was obtained from a commercial source (NanoLN, Jinan Jingzheng Electronics Co., Ltd.) in the form of a 75 mm wafer separated from a Si handle (0.4~mm thickness) by a 1.8 $\mu$m layer of SiO$_2$ (known as LN-on-insulator, or LNOI).  Dies were segmented and poled, then etched to form waveguides. Details of the poling process are reported in Ref.~\citen{Zhao2019}. We used a diagnostic method, described in Ref.~\citen{Ruesing2019}, to study and improve the poling process. After poling, the waveguide structure fabricated in collaboration with the University of Rochester, as reported in Ref.~\citen{Luo2018}. The upper cladding of the waveguide was left as air after etching.  

We designed waveguides with a length of 0.7~cm (poled section of length 0.5~cm) as shown in Fig.~\ref{fig-1}(a), and a ridge etch depth of 50~nm, with an angle of about 75$^\circ$ to the horizontal, as indicated in Fig.~\ref{fig-1}(b). The quasi-TE-polarized fundamental modes, shown in Fig.~\ref{fig-1}(b) were calculated using vectorial mode simulation software to have $A_\text{eff}=1.1\ \mathrm{\mu m}^2$ at 1570 nm wavelength, and $A_\text{eff}=0.4\ \mathrm{\mu m}^2$ at 785 nm wavelength. The waveguide is single-mode at the longer wavelengths, which is important for generating down-converted photons in spatially pure states and heralding. The normalized mode overlap integral between these modes was calculated to be 78\%. Figure~\ref{fig-1}(c) shows the effective refractive indices calculated for the relevant waveguide modes. Here, type-0 quasi-phase matching (QPM), i.e.,~matching of the effective refractive indices of the lowest-order TE-polarized waveguide modes at 1570~nm and 785~nm [labeled `TE0' in Fig.~\ref{fig-1}(c)], was achieved by poling. The calculated QPM period, $\Lambda$ = 2.8 $\mathrm{\mu m}$ and a first-order QPM grating was used. In this configuration, the nonlinear coefficient is $d_{33} = 27\ \text{pm}.\text{V}^{-1}$. Gold electrodes (with a thin chrome adhesion layer) were lithographically patterned and used for poling by applying voltage pulses. 
   
Evidence that the QPM grating was well fabricated was obtained from a conventional waveguide second harmonic generation (SHG) measurement, using as input a tunable-wavelength, continuous-wave pump around 1569 nm. While a detailed study of SHG will be presented elsewhere, here, we summarize the key observations. The converted power around 784.5 nm was measured when tuning the wavelength of the pump around 1569~nm. By fitting the central lobe of the measured SHG spectrum [see Fig.~\ref{fig-1}(d)] to a sinc-squared functional form, we measured a SHG conversion efficiency of nearly $2 \times 10^3$ \%.$\text{W}^{-1}$.$\text{cm}^{-2}$, with a full-width at half-maximum (FWHM) spectral bandwidth of 1.25~nm. This relatively narrow bandwidth (e.g., compared to Refs.~\citen{Chang2016,Wang2018b}) is suitable for telecommunications-band SPDC in LNOI waveguides, since the bandwidth is approximately that of a typical standard wavelength-division multiplexing (WDM) filter. In our fabricated waveguides, the poling period was not chirped, and therefore, sidelobes were present in the spectrum, but were separated from the main peak by several nanometers. Thus, any photons generated by these weak satellite peaks were filtered out by the filter. 

\section{{SPDC} Experimental Details}
\label{sec-experiment} 
Energy conservation between the pump ($P$) and the generated signal ($S$) and idler ($I$) photon pair dictates the relationship, $\omega_P = \omega_{S} + \omega_{I}$ between the optical frequencies, in radians, at the three respective wavelengths. Since one photon from the pump generates one each at the signal and idler wavelengths, the rates (units: $\text{s}^{-1}$ or Hz) of generated signal and idler photons, and the coincidence counts, are all expected to be linearly proportional to the pump power, as was experimentally verified.  

\begin{figure}[thb]
\centering
\includegraphics[width=\linewidth, clip=true]{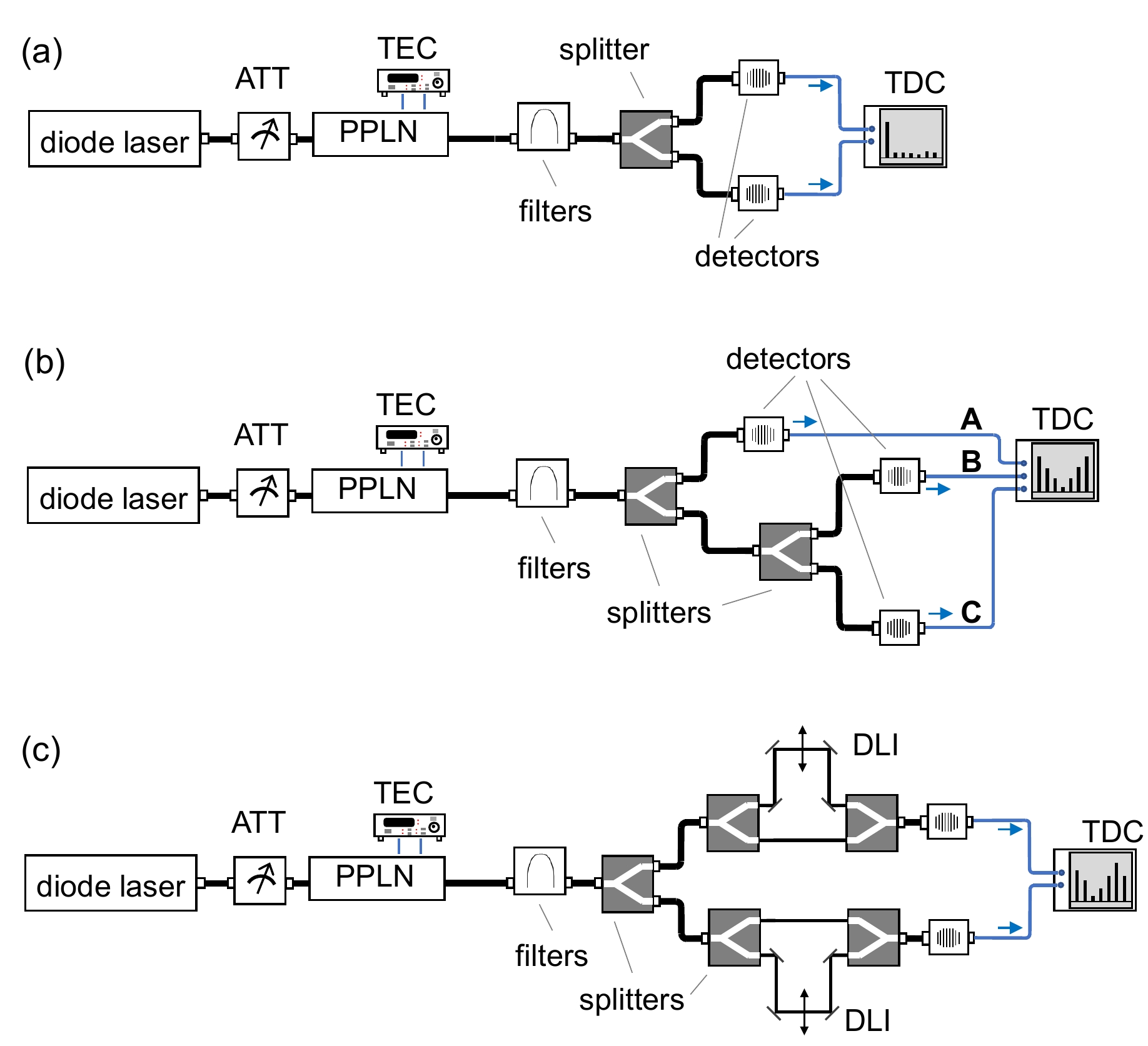}
\caption{(a) Schematic of the experiment to characterize frequency-degenerate photon pair generation, and measure the coincidences-to-accidentals ratio (CAR). (b) Schematic of the experiment to measure heralded single photon generation. (c) Schematic of the experiment to measure energy-time entanglement using a Franson interferometer. ATT: Variable attenuator. TEC: Thermo-electric controller. PPLN: periodically-poled lithium niobate waveguide. TDC: time to digital converter.} 
\label{fig-2}
\end{figure}
Measurements reported here used the experimental configurations for signal-idler cross-correlation and second-order heralded self-correlation shown in Fig.~\ref{fig-2}. The bare-die chip was mounted on a temperature-controlled stage with a thermo-electric controller (TEC) in feedback with a thermistor on the stage mount. The chip was maintained at a temperature of 63 $^\circ$C in order to tune the QPM peak [see Fig.~\ref{fig-1}(d)] to the input laser wavelength. Light was coupled to and from the chip using lensed tapered polarization-maintaining fibers designed for 1550 nm (Oz Optics Ltd.). The propagation loss at both wavelengths was estimated to be less than -1 dB/cm around 1570 nm, and less than -3 dB/cm at 785 nm, based on previous measurements.\cite{Luo2018} The chip was simply diced for measurement, without polishing, and thus incurred high, un-optimized coupling losses of -5 dB/facet at 1570 nm and -15 dB/facet at 785 nm. Loss values of -3 dB/facet at 1570 nm and -6.6 dB/facet at 785 nm can be simply attributed to mode-overlap mismatch between the tapered fiber and the waveguide modes as calculated by electromagnetic simulation software (Lumerical, Inc.), with the rest attributed to excess loss due to roughness or launch at 785 nm into an higher-order waveguide mode that is not phase-matched (the diced edge was quite rough in this chip). With the high intrinsic SHG conversion efficiency obtained here, the more important loss and coupling efficiency numbers for SPDC are those at the generated signal and idler wavelengths (i.e., near 1570~nm). Nevertheless, further improvements will be necessary before the overall performance (e.g., for heralding) can be comparable to traditional PPLN waveguides.   

Output light from the chip was routed through a cascade of two filters, where the first was a pump-reject filter, consisting of a tabletop assembly of a long-pass, free-space filter with two fiber collimators, and the second was a fiber-coupled, off-the-shelf, telecommunications-grade optical filter centered around 1568.9 nm with bandwidth 0.8 nm. The insertion loss of the filter cascade at the wavelengths of interest was less than -4~dB and the pump rejection should be greater than 150~dB. The filter assembly was followed by a 50\%-50\% splitter. For frequency-degenerate photon pairs incident on the splitter, both photons end up in the same detector one-half of the time, and would not be counted as a coincidence by the TDC. An imperfect splitting ratio resulted in a slight difference in the measured singles rates of the signal and idler channels. The spectral width of the QPM curve was approximately similar to the pass-band width of the filter; thus, filtering does not reduce the flux rate or brightness of the photon pairs by much.  

Photons were detected using fiber-coupled superconducting nanowire single photon detectors (SNSPD), cooled to 0.8~K in a closed-cycle Helium-4 cryostat equipped with a sorption stage (Photon Spot, Inc.). The detection efficiencies of the two SNSPD's used for coincidence measurements were about 68\% and, for one of the detectors used in heralding experiments as described below, was about 90\%. The detectors were not gated and were operated in a simple dc-biased mode with an RF-amplified readout. Detected signals were processed using a time-to-digital converter (TDC) instrument (qutools GmbH), with coincidence window of 15~ns or 10~ns. Singles and coincidences due to dark counts were measured separately, but since their contribution was seen to be negligible, they were not subtracted from the measurements. Each histogram peak was fitted by a Gaussian function, whose FWHM was measured to be typically 27~ps. 

The SPDC process was pumped using a continuous-wave laser; as such, the joint spectral intensity of the two-photon state is not factorizable. In this preliminary study, waveguides of only one length (5 mm poled region length) were used; the optimum waveguide length and pulse width of the pump both play a role, along with the group velocity at the pump and down-converted wavelengths, in determining the optimal factorability condition.\cite{inbook-Silberhorn}  

\section{Measurements}
Typical characterization measurements for SPDC report pair generation rates (PGR) and two-photon correlation measurements, quantifying the pair generation and heralding properties of the source. Since the purpose of this paper is to demonstrate that high-quality pairs can be experimentally generated using LNOI waveguides, we focus mainly on the low pump-power case; nevertheless, an appreciable rate of pairs and single photons was measured because of the high brightness of the source. 

\subsection{Pair Generation Rate (PGR) and Brightness}
\label{subsec-counts}

\begin{figure}[thb]
\centering
\includegraphics[width=\linewidth, clip=true]{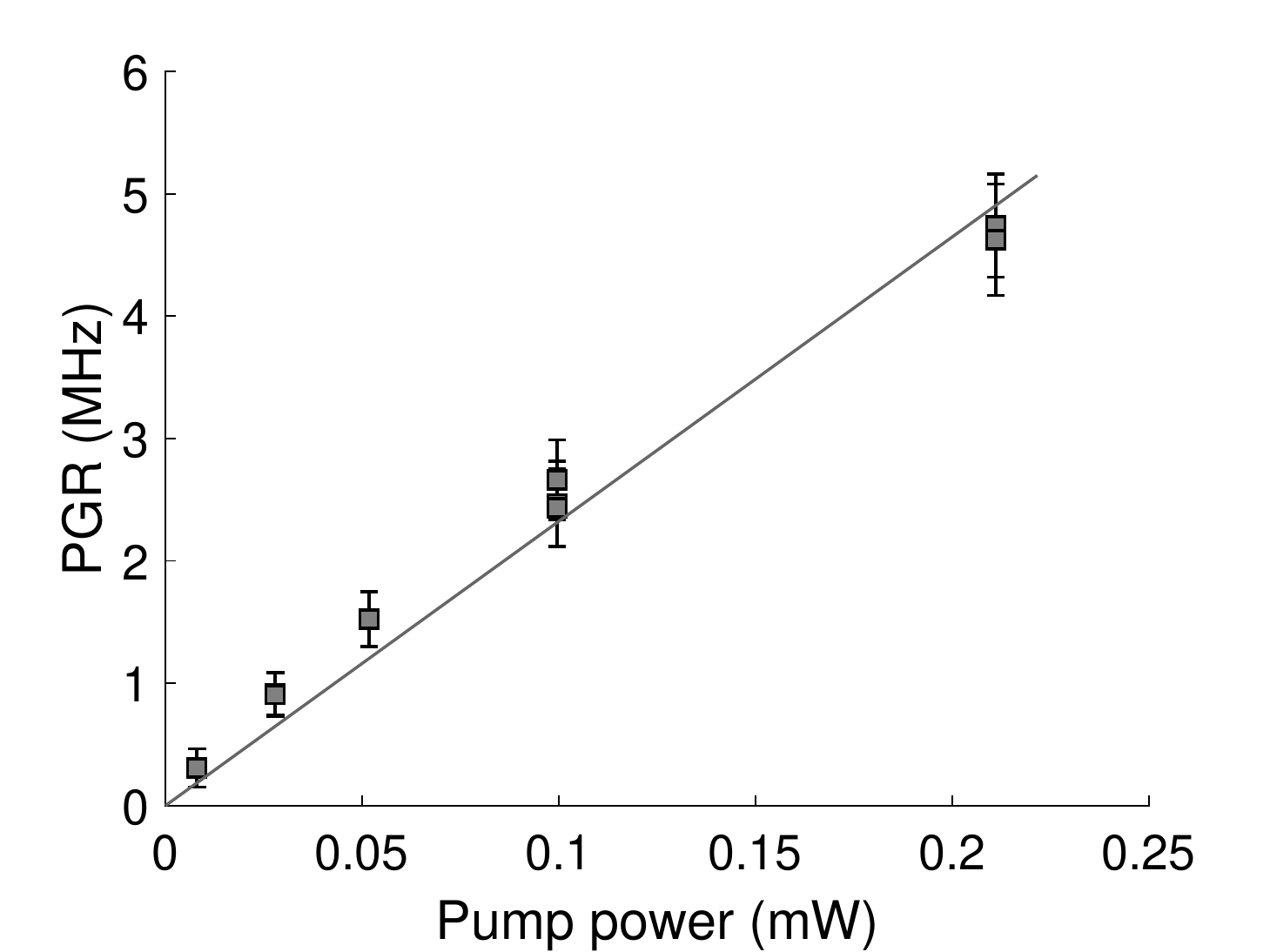}
\caption{Pair generation rate (PGR, units: MHz), also known as the on-chip pair flux, versus pump power (units: mW) in the waveguide.} 
\label{fig-3}
\end{figure}

The on-chip pair flux, also known as the pair generation rate (PGR), is calculated as $\text{PGR} = (N_S N_I)/(2 N_{SI})$, where $N_{S,I}$ are the measured singles rates of the signal and idler photon detection events, respectively, and $N_{SI}$ is the rate of the signal-idler coincidence detection events. The factor of $2$ in the denominator is because of the beam splitter in the signal path of the frequency-degenerate photons, which results in one-half of the generated pairs, at best, being counted. In the other cases, the signal and idler photons are incident at the same detector. We calculated PGR by averaging the time-resolved traces of the counts, for several different values of the input pump power, as shown in Fig.~\ref{fig-3}. The slope of the fitted line is $23\ \text{MHz}.\text{mW}^{-1}$. There are many reports of SPDC using traditional PPLN waveguides; a typical state-of-the-art value for the slope efficiency of PGR with pump power is about $14\ \mathrm{MHz}.\mathrm{mW}^{-1}$.\cite{PhysRevLett.113.103601}

Dividing further by the FWHM of the filter used before the single-photon detectors, $\Delta \lambda = 0.8$~nm (97~GHz), we calculate the brightness of our pair source to be $B=2.9 \times 10^7\ \mathrm{pairs}.\mathrm{s}^{-1}.\mathrm{nm}^{-1}.\mathrm{mW}^{-1}$, or $B=3 \times 10^5\ \mathrm{pairs}.\mathrm{s}^{-1}.\mathrm{GHz}^{-1}.\mathrm{mW}^{-1}$, depending on the units used for the bandwidth. This is of the same order-of-magnitude as Ref.~\citen{Chen2019a_YPH_arxiv} ($7 \times 10^7\ \mathrm{pairs}.\mathrm{s}^{-1}.\mathrm{nm}^{-1}.\mathrm{mW}^{-1}$), which also used a periodically-poled LNOI waveguide, and the traditional titanium-indiffused PPLN waveguide of Ref.~\citen{Montaut2017} ($1.4 \times 10^7\ \mathrm{pairs}.\mathrm{s}^{-1}.\mathrm{nm}^{-1}.\mathrm{mW}^{-1}$).

\subsection{Coincidences-to-Accidentals Ratio (CAR)}
\label{subsec-CAR}
\begin{figure}[thb]
\centering
\includegraphics[width=\linewidth, clip=true]{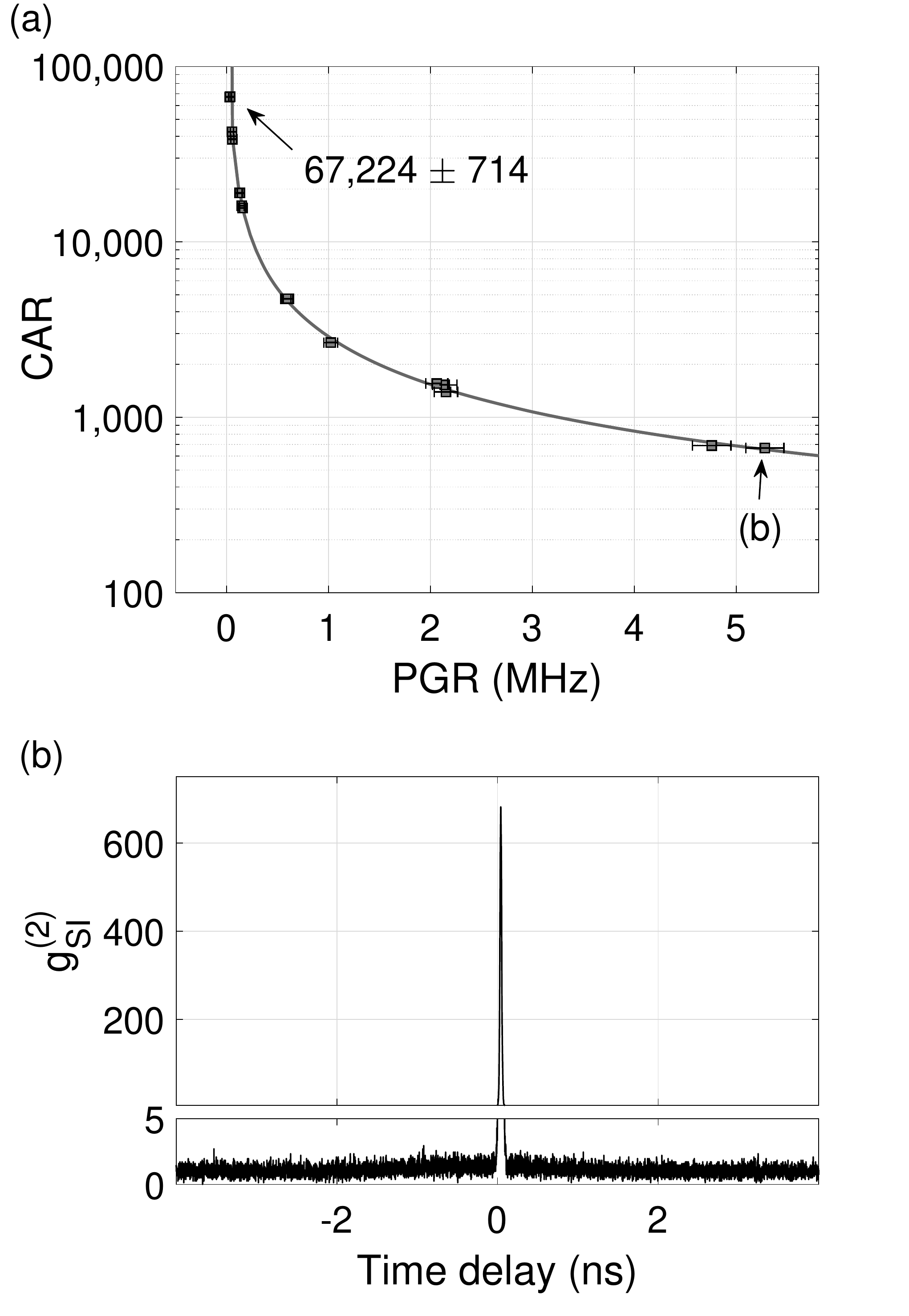}
\caption{(a) Coincidences-to-Accidentals Ratio (CAR) versus the pair generation rate, PGR. The highest measured value is indicated. The error bars are one standard deviation in each direction (vertical error bars are too small to visualize at this scale). (b) The signal-idler cross-correlation counting histogram for the lowest CAR value. The full-width at half-maximum of the central peak was 27~ps.} 
\label{fig-4}
\end{figure}

Achieving high CAR depends on low detector noise, suppression of pump and scattering noise, improvement of the stability of pair generation, and improving factors such as loss in the device and the experimental setup, which lead to broken pairs and increase the rate of accidental coincidences. Figure~\ref{fig-4}(a) shows the measurements of the CAR versus the (on-chip) coincidence rate. The CAR was calculated as $\text{CAR} = \text{max} [g^{(2)}_\text{SI}(t)]-1$ from the normalized signal-idler cross-correlation, $g^{(2)}_\text{SI}(t)$, which was obtained from the histogram of signal-idler coincidences that was measured by the TDC instrument as a function of the delay $t$ between the two channels. The histograms were acquired in start-stop mode for a measurement time $T$ that varied between 120~s and 480~s (at lower power), and a coincidence window $W = 15 \ \text{ns}$.  It was verified that each coincidence peak was well fit by a Gaussian function, whose FWHM was $25.2 \pm 0.4$~ps across all the datasets reported in Fig.~\ref{fig-4}(a). The value of CAR reported here is obtained from the amplitude of the fitted peak, which is slightly less than the raw measured value of $ \text{max} [g^{(2)}_\text{SI}(t)]$. 

The highest CAR was $67,224 \pm 714$ measured when the PGR was $8.2 \pm 5.7 \times 10^4\ \text{pairs}.\text{s}^{-1}$ (detected pair flux: $24 \pm 17\ \text{pairs}.\text{s}^{-1}$). At the highest power values used here, CAR = $668 \pm 1.7$, at PGR = $1.2 \pm 0.08 \times 10^7\ \text{pairs}.\text{s}^{-1}$ (detected pair flux: $3.6 \pm 0.3 \times 10^3\ \text{pairs}.\text{s}^{-1}$). CAR decreased at higher pump powers and thus, at higher PGR, as expected, following the trend line $\text{CAR} \propto \text{PGR}^{-1}$, as shown in Fig.~\ref{fig-4}. 

For comparison with previous reports of SPDC in LNOI structures, maximum CAR values of about 6-15 in waveguides~\cite{Rao2018a}, up to 43 in microdisk resonators,\cite{Luo2017} and up to 600 in waveguides (at PGR = $0.8 \times 10^6\  \text{pairs}.\text{s}^{-1}$)~\cite{Chen2019a_YPH_arxiv} were reported. A comparison with our LNOI waveguides is provided in Table~\ref{tab-SPDC-LNOI}. 

For comparison with traditional PPLN waveguides, at high pair flux, a value of CAR = $72$ at PGR = $3.9 \times 10^7\ \text{pairs}.\text{s}^{-1}$ was reported in Ref.~\citen{Montaut2017}; thus, the product of CAR and PGR is about 3 times higher in our LNOI waveguides. At low pump powers, CAR values in the thousands have been measured using reverse-proton-exchange PPLN waveguides; for example, Ref.~\citen{Zhang2007} reports CAR = $4,452$ at PGR = 2~MHz using a pulsed source (and 60~ps coincidence time window), which is approximately the same as achieved here using a continuous-wave source (and much longer coincidence time window, 15~ns). The highest reported value (to our knowledge) of CAR is $8\times 10^5$ measured, however, at PGR of only 5 $\text{pairs}.\text{s}^{-1}$;~\cite{Inagaki:13} in comparison, the product of CAR and PGR is about four orders-of-magnitude higher in our LNOI waveguides. 

\begin{figure}[thb]
\centering
\includegraphics[width=\linewidth, clip=true]{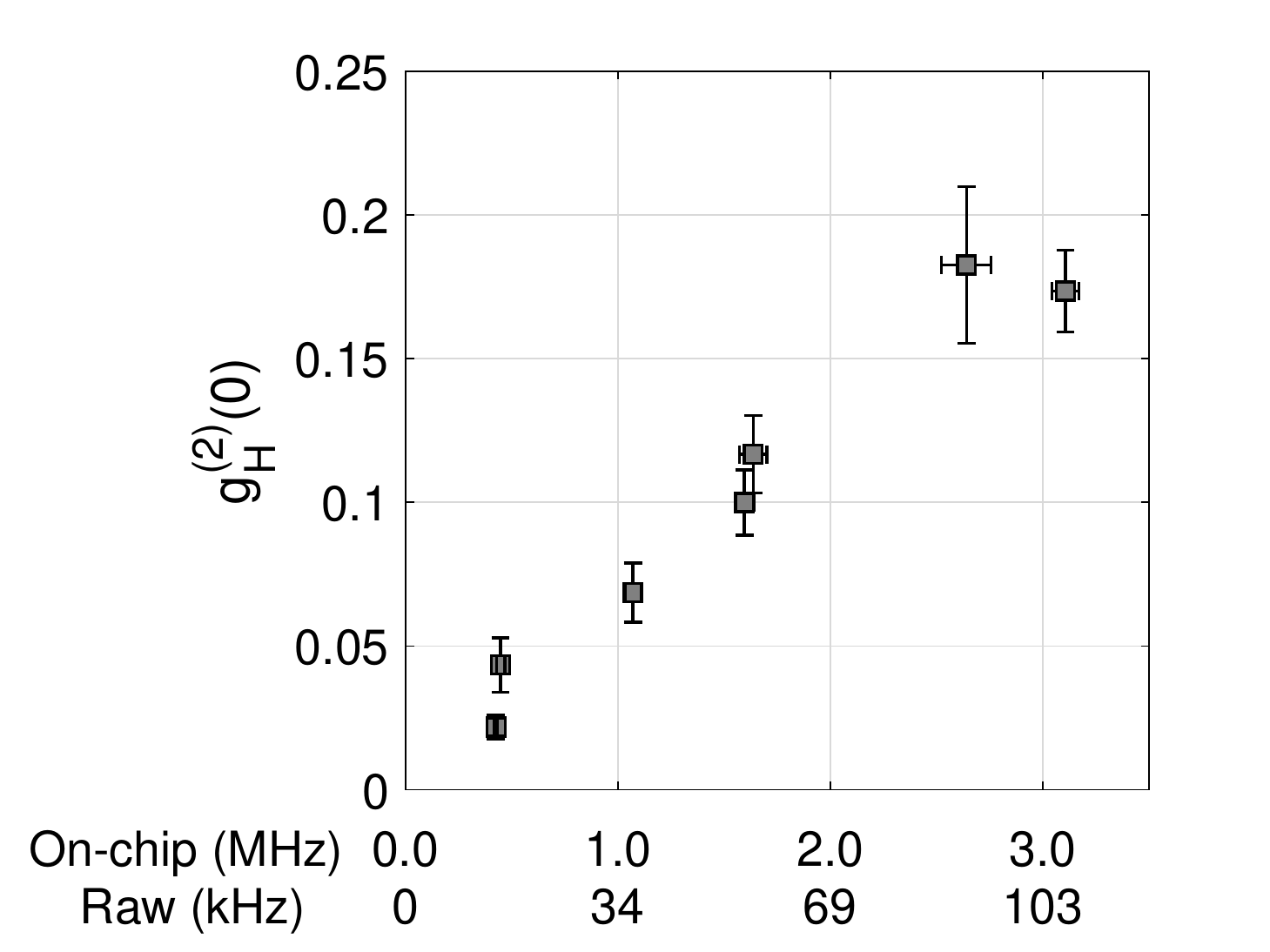}
\caption{Heralded single photon generation. Conditional self-correlation (heralded auto-correlation) $g^{(2)}_H(0)$ measured using the setup shown in Fig.~\ref{fig-2}(b). The horizontal axis shows both the raw herald rate (kHz) and on-chip singles rate (MHz); the latter is obtained by dividing by the measured losses between the chip and the detector. The error bars are one standard deviation. The lowest measured  $g^{(2)}_H(0)$ was $0.022\, \pm\, 0.004$.} 
\label{fig-5}
\end{figure}

\subsection{Heralded single-photon generation}
\label{subsec-g2}
Detecting one photon of the pair results in a heralded single-photon source, since the other photon is expected to show non-classical anti-bunching behavior. This waveguide supports a single propagating mode at the down-converted wavelengths near $1.57\ \mathrm{\mu m}$ wavelengths, and, after filtering, the central QPM peak is fairly narrow, as shown in Fig.~\ref{fig-1}(d). Therefore, in contrast to SPDC in bulk crystals, the signal and idler photons are emitted into a pair of discrete, well-defined modes which propagate collinearly. Here, as stated earlier, the width of the QPM peak is approximately the same as that of the passband of the filter for the signal and idler photons. We have not yet measured the two-photon joint spectrum, and thus cannot conclude that the heralded photon is actually in a pure single-photon Fock state. Accordingly, the discussion for the present is restricted to the measurement of the second-order correlation function of the heralded photon, i.e.,~a characterization of anti-bunching.  

Figure~\ref{fig-5} shows the heralded (i.e.,~conditional) single-photon second-order self-correlation function, $g^{(2)}_H(0)$, obtained by detecting one of the generated photon pair as a herald, and measuring the self-correlation of the other photon in the presence of the herald. The normalized value of the photon correlation measurement on the heralded single photons at zero time delay was calculated using the formula~\cite{Beck2007} $g^{(2)}_H(0)=N_{ABC}N_{A}.\left(2 N_{AB}N_{AC}\right)^{-1}$, where $N_{A}$ is the average photon detection rate on the heralding SNSPD detector [labels are shown in Fig.~\ref{fig-3}(b)], double coincidences $N_{AB}$ and $N_{AC}$ correspond to average rates of simultaneous events on one of the detectors (B or C) and the heralding SNSPD detector (A), and triple coincidences $N_{ABC}$ correspond to average rates of simultaneous events on all three detectors. This parameter has also been called the anti-correlation parameter.\cite{Grangier1986} Note that $g^{(2)}_H(0)$ can also be written in terms of the probability of observing a single photon in the signal arm and the probability of observing two photons in the signal arm, $g^{(2)}_H(0)=2 N_{A}N_{ABC}.\left(N_{AB}+N_{AC}\right)^{-2}$, where the factor of $2$ in the numerator comes from the splitting ratio of the beamsplitter.\cite{URen2005} 

Double and triple coincidences were defined as simultaneous detections within a 10~ns time window, measured directly by the TDC hardware (calculating coincidences between combinations of input channels without software post-selection). Counting times were 200 seconds (with 1000 seconds for one point as a check). 

Even at the highest power values used in this sequence of measurements, $g^{(2)}_H(0) = 0.183 \pm 0.03$, well below the classical threshold, at an on-chip (i.e., inferred) heralding rate of $N_{A}=3.1$~MHz (raw measured herald rate 107 kHz). At lower pump powers, values as low as $g^{(2)}_H(0) = 0.022 \pm 0.004$ were directly measured (the errorbar is one standard deviation uncertainty), for an on-chip heralding rate of $N_{A} = 0.43$~MHz (raw measured herald rate 15 kHz). There have not been previous reports of heralded single-photon generation using LNOI waveguides or resonant devices. For comparison with traditional PPLN devices, $g^{(2)}_H(0) = 0.023$ has been measured at (detected) $N_{A} = 2.1$~MHz,\cite{LPOR:LPOR201400404} and $g^{(2)}_H(0) = 0.005$ has been measured at (detected) $N_{A} = 10$~kHz.\cite{art-Brida-2012}  

The heralding (Klyshko) efficiency, defined as $N_{AB}/(N_{A} \times D)$ where $D$ is the detection efficiency of the heralded photon, was calculated to be between 1.3\% and 2\% for raw (off-chip) values of the rates. The main reason for the low efficiency of the off-chip efficiency is loss: the sum of the fiber-to-waveguide loss and the insertion loss of the filters is about -14 dB for each of the SPDC-generated photons in the current device and experimental configuration, whereas, for example, the coupling efficiency to fiber in Ref.~\citen{LPOR:LPOR201400404} was estimated to be about 60\% and the equivalent transmission efficiency through free space was above 90\% in Ref.~\citen{Krapick2013}. Since the ratio of the on-chip pair rate to the on-chip singles rate (averaged between the signal and the idler) exceeds 52\%, we expect heralding efficiencies comparable to established PPLN technology upon improvements in the coupling from these LNOI waveguides to detectors through fibers or free space.      

\begin{figure*}[thb]
\includegraphics[width=0.7\linewidth, clip=true]{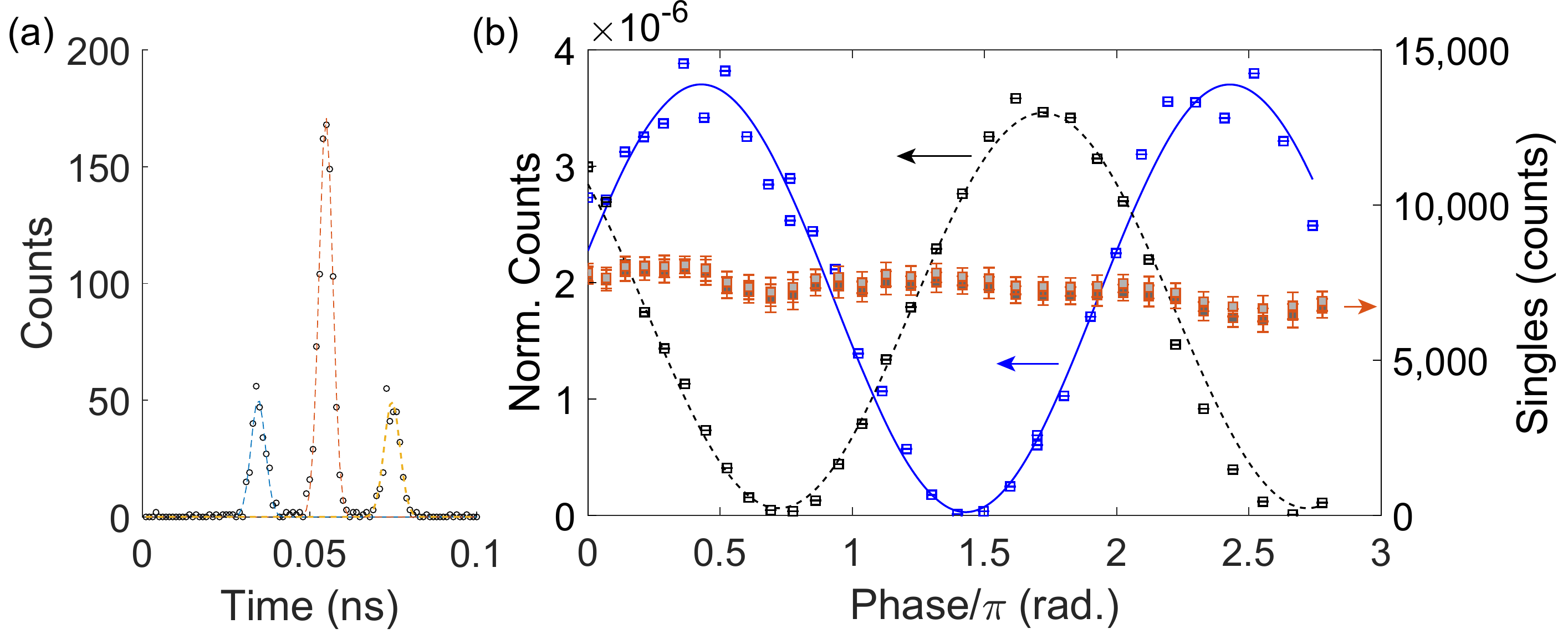}
\caption{(a) Representative histogram for the measurement of energy-time entanglement, at a particular phase setting of the delay line interferometers (DLI's).  (b) Two-photon interference pattern measured as the phase of one DLI is swept. The interference pattern for two different phase settings on the second DLI are shown. Black and blue dots (with errorbars): experimental data, black solid and dashed lines: sinusoidal fit. The right-hand side axis shows the singles counts averaged over the acquisition time, measured at the same time as the two-photon coincidences. } 
\label{fig-6}
\end{figure*}

\subsection{Energy-time entanglement}
\label{subsec-Entanglement}
The generated photon pair is expected to demonstrate energy-time entanglement which can be investigated through a Franson-type two-photon interference experiment, by violating Bell's inequality~\cite{franson1989bell,kwiat1993high}. Figure~\ref{fig-6} shows the measurement of visibility fringes using an unfolded Franson interferometer configuration, whose schematic is shown in Fig.~\ref{fig-2}(c).

Two fiber-coupled, polarization-maintaining, piezo-controlled delay-line interferometers (DLI's), each with an FSR of 2.5~GHz and peak-to-valley extinction ratio approximately 25 dB were used in these measurements. Data scatter in the fringes was caused by fluctuations of photon pair flux coupled to the DLI's, which was mainly a result of a drift in the state of polarization or variations in fiber-to-chip alignment. However, this effect is minor and negligible compared to the expected variations, with relative phase, in the two-photon coincidence counts. The raw measured coincidences showed three peaks, as shown in typical data plotted in Fig.~\ref{fig-6}(a). In each case, a Gaussian function was seen to be a good fit to the raw data, and the peaks were clearly separated, leading to a simple and robust fit. The fitting uncertainty (one standard deviation) is shown as the errorbar in the plotted points and is too small to be visible. The phase of one of the DLI's (i.e., the phase delay between the short arm and the long arm of that DLI) was held constant at two different settings, and the phase of the other DLI was swept over approximately one free spectral range, tuned by voltage. In Fig.~\ref{fig-6}(b), the normalized coincidence counts were calculated by dividing the raw measured counts (which varied from about 0.37 to 189 counts in the measurement time) by the product of the singles counts during the same time; this normalized quantity factors out the minor variations in the singles counts with time also shown in Fig.~\ref{fig-6}(b). The flat singles rates (versus phase) show the absence of single-photon interference, as desired.

Proof of photon pair entanglement requires a two-photon interference pattern fringe visibility $V \geq 70.7\%$ (without necessarily providing a test of local realism)\cite{Clauser1974}. The fitted measurements showed $V$ clearly in excess of this threshold value, measured when the on-chip PGR was about 235~kHz, as inferred from the recorded singles rates and the coupling losses. Previously, a CAR measurement was performed at a similar PGR, and this is recorded in Table~\ref{tab-SPDC-LNOI} alongside the $V$ values. From the raw data, i.e.,~the highest and lowest value in Fig.~\ref{fig-6}(b), we calculated $V_\text{data} = 99.3 \pm 1.9 \%$ (data points shown in blue) and $V_\text{data} = 99.5 \pm 1.8 \%$ (data points shown in black) for the two phase settings of the unfolded Franson configuration. The indicated errorbar is the uncertainty which arises from the goodness-of-fit of the parameters of the Gaussian function used to fit the central peak; in many cases, the size of the errorbar is too small to be visible. From a fit to the entire ensemble of measurements based on the non-linear least-square curve fitting algorithm in Matlab, we obtained $V_\text{fit} = 98.4 \%$ and $V_\text{fit} = 96.4 \%$ for the two cases. These measurements confirmed the energy-time entanglement properties of the pairs, as shown by the sinusoidal variation of coincidences with phase. 

\begin{table*}
\caption{Recent results of entangled photon-pair and heralded single-photon generation near $1.55\ \mathrm{\mu m}$ wavelengths using optically-pumped SPDC in LNOI photonic devices.}
\label{tab-SPDC-LNOI}
\raggedleft
\begin{tabular}{|c|c|c|c|c|c|} \hline\hline
Ref. & Structure & PGR & CAR & $g^{(2)}(0)$ & Visibility \\ \hline\hline
Main et al.\cite{Main2016} 		& waveguide & [theory]  & - & - & - \\
Frank et al.\cite{Frank2016} 		& microdisk & 450~kHz  & 6 $^{(a)}$ & - & - \\
Luo et al.\cite{Luo2017}				& microdisk & 0.5 Hz $^{(b)}$ & 43 & - & -\\
Rao et al.\cite{Rao2018a} 			& waveguide & 7 kHz $^{(c)}$ & 15 & - & -\\
''& ''& 28 kHz $^{(c)}$ & 6 & - & - \\
Chen et al.\cite{Chen2019a_YPH_arxiv} 			& waveguide & 0.8 MHz & $631 \pm 210$ & - & - \\ \hline
This work							& waveguide & 	82 kHz & $67,224 \pm 714$ & - & -\\
'' & '' & 235 kHz$^{(d)}$ & $16,075 \pm 87^{(e)}$  & - & $\bigg\{$\begin{tabular}{c}$99.3 \pm 1.9 \%$\\ $99.5 \pm 1.8 \%$ \end{tabular}\\
'' & '' & 430 kHz &6,250 $^{(f)}$ & $0.022 \pm 0.004$ & -\\
'' & '' & 3.1 MHz & 1,043 $^{(f)}$ & $0.183 \pm 0.03$ & -\\
'' & '' & 12 MHz & $668\pm 1.7$ & - & -\\ 
[1ex] \hline
\end{tabular}
\newline 
\begin{minipage}{\textwidth}
$^{(a)}$ Estimated from the peak-to-side-lobe ($\pm 0.5$~ns) ratio of coincidence counts. $^{(b)}$ Peak value of the raw coincidence counts divided by the measurement time, and further divided by the detection efficiency and detector gating duty cycle. \, $^{(c)}$ From the stated pump power and on-chip pair generation rate.  $^{(d)}$ Estimated from the measured singles rate. $^{(e)}$ From a separate measurement of CAR at the PGR for the visibility measurement. $^{(f)}$ From the fitted line in Fig.~\ref{fig-3}.
\end{minipage}
\end{table*}

\section{Conclusion}
\label{sec-discussion}
In conclusion, we have demonstrated high-quality photon-pair and heralded single-photon generation at telecommunications wavelengths using a periodically-poled thin-film lithium niobate waveguide. Small mode cross-sectional area waveguides exhibit a higher refractive index dispersion, and are more sensitive to fabrication imperfections than traditional LN waveguides, and the poling process for LNOI waveguides has only been relatively recently studied. Non-destructive, diagnostic images of the poling outcomes were used here to identify which sections of the test chip, which contained several different poled regions, were suitable for waveguide formation and measurement. In practice, the waveguide ridge width and etching depth could be tailored to the measured properties of the poling process. As a measure of the quality of the photon pairs, we measured the highest CAR yet reported, by far, for thin-film LN waveguides, $67,224 \pm 714$ measured when the PGR was $8.2 \pm 5.7 \times 10^4\ \text{pairs}.\text{s}^{-1}$. 

Table~\ref{tab-SPDC-LNOI} summarizes the significant progress thus achieved in entangled pair generation and heralded single photon generation using SPDC in thin-film LNOI photonic structures. In fact, our waveguides achieve high CAR at high PGR, with high two-photon interferometric visibility, similar to the best traditional PPLN waveguides. However, compared to the latter, the off-chip coupling losses of our microchips are substantially higher at the present time, and the chips are not packaged and fiber pigtailed; thus, the overall maturity of this integrated thin-film device is not yet as advanced as the highly optimized and engineered ``plug and play'' traditional LN SPDC device.\cite{Montaut2017} Therefore, although the values of $g^{(2)}_H(0)$ are low, showing the high quality of heralded single photon generation, the heralding rates are also low, as is the Klyshko efficiency. The ratio of on-chip pairs-to-singles rate is higher than 50\%, suggesting, on the one hand, that better fiber coupling and packaging will lead to a high-quality and useful stand-alone device, and on the other hand, that the present device structure is already suitable for on-chip integrated quantum photonics. In this report, we have established the intrinsic properties of periodically-poled thin-film MgO:LN waveguides with sub-micron modal area as suitable candidates for high-quality and efficient SPDC, and the next steps are to improve out-coupling and packaging properties in order to benefit heralding and practical applications, where traditional PPLN waveguides still continue to dominate, and further develop the technology of on-chip integration. 

\section*{Funding}
National Science Foundation (NSF) (ECCS 1201308, EFMA-1640968 ``ACQUIRE: Microchip Photonic Devices for Quantum Communication over Fiber''). 

\section*{Acknowledgments}
The authors are grateful to Prof. Q. Lin, U. Javid, J. Ling and M. Li (University of Rochester) for collaboration on waveguide fabrication.




\end{document}